\pdfoutput=1
\documentclass[copyright,creativecommons]{eptcs}
\providecommand{\event}{FBTC 2010} 
\providecommand{\volume}{??}
\providecommand{\anno}{2010}
\providecommand{\firstpage}{1}
\providecommand{\eid}{??}

\usepackage{amssymb,amsmath}
\usepackage[latin1]{inputenc}
\usepackage[pdftex]{graphicx}
\usepackage{xspace}
\usepackage{fancybox}

\newcommand{\james}{JAMES II\xspace}
\newcommand{\famval}{FAMVal\xspace}
\newcommand{\fada}{FADA\xspace}

\newcommand{\eg}{e.g.,\xspace}
\newcommand{\ie}{i.e.,\xspace}

\title{A flexible architecture for modeling and simulation of diffusional association}
\author{Fiete Haack
\institute{University of Rostock}
\institute{Institute of Computer Science\\
Albert-Einstein-Str. 21\\
18059 Rostock, Germany}
\email{fiete.haack@uni-rostock.de}
\and
Stefan Leye
\institute{University of Rostock}
\institute{Institute of Computer Science\\
Albert-Einstein-Str. 21\\
18059 Rostock, Germany}
\email{stefan.leye@uni-rostock.de}
\and
Adelinde M. Uhrmacher
\institute{University of Rostock}
\institute{Institute of Computer Science\\
Albert-Einstein-Str. 21\\
18059 Rostock, Germany}
\email{adelinde.uhrmacher@uni-rostock.de}
}
\def\titlerunning{Flexible Architecture for Diffusional Association}
\def\authorrunning{Fiete Haack \& Stefan Leye \& Adelinde M. Uhrmacher}
\begin{document}
\maketitle

\begin{abstract}
%
%


Up to now, it is not possible to obtain analytical solutions for complex molecular association processes (e.g. Molecule recognition in Signaling or catalysis). 
Instead Brownian Dynamics (BD) simulations are commonly used to estimate the rate of diffusional association, e.g. to be later used in mesoscopic simulations. \\
Meanwhile a portfolio of diffusional association (DA) methods have been developed that exploit BD. 
However, DA methods do not clearly distinguish between modeling, simulation, and experiment settings. 
This hampers to classify and compare the existing methods with respect to, for instance model assumptions, simulation approximations or specific optimization strategies for steering the computation of trajectories. \\
To address this deficiency we propose FADA (Flexible Architecture for Diffusional Association) - an architecture that allows the flexible definition of the experiment comprising a formal description of the model in SpacePi, different simulators, as well as validation and analysis methods.   \\
Based on the NAM (Northrup-Allison-McCammon) method, which forms the basis of many existing DA methods, we illustrate the structure and functioning of FADA.
A discussion of future validation experiments illuminates how the FADA can be exploited in order to estimate reaction rates and how validation techniques may be applied to validate additional features of the model.

\end{abstract}

\section{Introduction}

Quantitative modeling and simulation in systems biology is often hampered by the lack of reaction rates, which require repetitive time resolved measurements in the wet lab.  
Wet lab experiments are extremely costly with respect to facilities and time and one estimated rate often corresponds to someone's doctoral degree \cite{Takahashi:2008}. 
Thus, it is not surprising that the idea of calculating reaction rates from "`first principles"' and hence to forego the need of wet-lab experiments is quite attractive. \\
Every reaction consists of a diffusion controlled binding event and the subsequent chemical reaction - accompanied with conformational changes. 
In case of diffusion-controlled reactions, \ie reactions in which the reaction process is faster than the diffusional encounter of the particles, the rate of association is the rate limiting step.
Therefore the estimation of the diffusional association (DA) rate is the first and most important step in determining the reaction rate of a chemical reaction. \\
A large portfolio of stochastic \cite{Northrup:1984, Wade:1993, Berry:2002} and continuum models (\cite{Chan:1984, Smart:1998}) has been developed and implemented in several software packages \cite{Briggs:1995, Gabdoulline:1997, Song:2004}. 
However, in these approaches model, simulator, and experiment are closely intertwined as they have been developed with an emphasis on execution efficiency rather than separation of concerns. 
As a consequence, existing methods can hardly be related or even compared to each other. 
This hampers reuse and extensions. 
It has been shown that the efficiency of simulators depend on the model \cite{Jeschke2008}. Consequently the development of different simulators for the same model and their configuration on demand is one source of efficiency \cite{Ewald:2009b}, which in the above approaches cannot be exploited easily. 
Furthermore, an impartial validation of the models is impossible, since the lacking separation between model and simulator does not allow experiments with different simulation engines in order to identify the valid ones.\\
The scope of this work is to present the software architecture \fada (Flexible Architecture for Diffusional Association), implemented in JAMES II (JAva-based Multipurpose Environment for Simulation) \cite{Himmelspach2007}. 
\fada unites different models and simulation engines for the computation of diffusional association processes and is further structured in terms of modeling, simulation and experiment settings. 
By exploiting the plug-in architecture of JAMES II, we can incorporate independent implementations of those components. 
Our aim is to include independent model components for the representation of particles, interparticle forces, and motion functions, which allows the flexible parametrization of the model for a repeatable and comparable execution in the context of an experiment. 
At the same time the architecture supports simulation engines, that are independent of the parametrization of the model. 
The simulation engines themselves may also contain independent components, like random number generators, parallelizations, etc., which can be exchanged or extended. 
\fada further allows experimentation with the created models, by exploiting existing methods as well as allowing the implementation of new methods for experiment design and analysis. \\
Based on the simple stochastic NAM (Northrup-Allison-McCammon) method \cite{Northrup:1984}, we present a formal description of a basic model for stochastic modeling of association processes.
Thereby, the SpacePi-Calculus \cite{John2008} serves as description language. 
Using the presented toymodel, we show an exemplary workflow for a validation experiment and discuss the most important steps of the validation experiment. 
Finally, we present first results and give an outlook to our future experiments. 

\section{Brownian Dynamics in Diffusional Association}

Bimolecular binding plays a central role in numerous biological processes, reaching from ligand binding, protein-protein encounter to signal transmission at synaptic junctions. 
The bimolecular binding process can be divided into two parts: the diffusional association to form an encounter complex, and the subsequent formation of the fully bound complex by nondiffusional rearrangements. 
The encounter complex is more considered to be an ensemble of possible conformations, than a well defined (transition) state \cite{Berg:1985}. 
Due to the relatively slow diffusion of macromolecules in solutions, the potential reaction partners stay in close vicinity once they met and thus undergo several microcollisions before diffusing apart. 
By this means the molecules' reactive groups can be properly aligned, such that subsequent conformational changes and eventually the desired chemical reaction may occur. 
As the rate of diffusional association marks an upper limit for the binding process, we consider the diffusional encounter of two particles, \ie the estimation of bimolecular association rates in the first place.\\
Brownian dynamics (BD) simulations are the most common approach to simulate the diffusional encounter. 
BD describes the diffusive motion of particles in solution, whose mass is significantly larger than the surrounding solvent. 
Thereby, the molecules may still be represented on an accurate atomic level (including interparticle forces), but the internal dynamics of the molecule are disregarded and the surrounding solvent is only implicitly taken into account by stochastic forces and friction. 
As a result of these simplifications the resolution of macromolecular motion is reduced to $\sim1~ ${\AA}. 
This allows BD to perform larger time steps ($\sim$1 picoseconds for proteins under normal viscosity conditions) compared to MD simulations, while the resolution is still detailed enough for the simulation of molecular binding processes. \\
In order to compute the actual rate of diffusional association, the Smoluchowski diffusion equation \cite{Smoluchowski:1917} needs to be solved.  
This equation describes the time evolution in terms of the probability density function of a particle undergoing Brownian movement in a force field.
Due to the large variety of simultaneous interactions, the diffusion equation can usually only be solved analytically for systems with simple geometry and rudimentary modelling of interparticle forces. 
This is why most DA-methods are based on the work of Northrup, Allison and McCammon \cite{Northrup:1984}, who established to connect BD simulations with the estimation of association rates.  \\
With increasing computational capabilities, there have also been successful attempts to use continuum models for the estimation of association rates, recently \cite{Song:2004}.
However, the accurate modeling of diffusion-controlled processes in the context of realistic biomolecular systems is still challenging. 
The incorporation of the effects of ionic strength, protein charges, and steric repulsions has proven to be the rate limiting step for most of the existing approaches.
Therefore, it is important to find simple but accurate model descriptions for these interparticle forces and effective algorithms for their evaluation. 
This applies for both approaches, continuum and particle based methods.


\subsection{State of the art}
\label{sec:stateoftheart}

\begin{figure}[htb!]
	\centering
		\includegraphics[width=9cm]{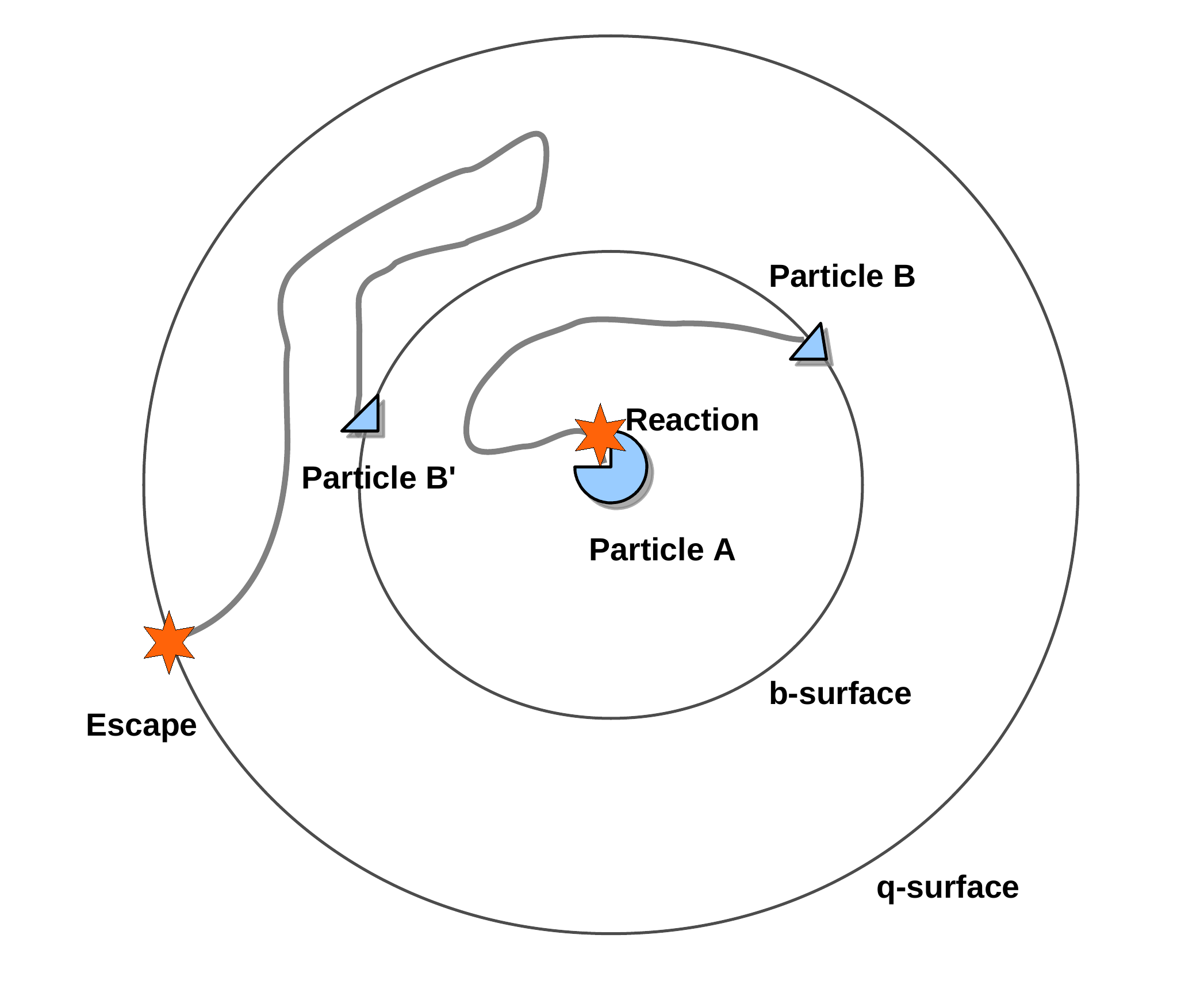}
\caption{Schematic representation of the NAM method. Particle B associates with particle A at the active site while Particle B` diffuses into infinite seperation.}
\label{fig:NAMmethod}
\end{figure}

One of the most common setup for the simulation of biomolecular diffusional association is incorporated in the NAM (Northrup-Allison-McCammon) method \cite{Northrup:1984}. 
The basic setup for the NAM method is rather simple, as illustrated in figure \ref{fig:NAMmethod}. 
We consider two particles of arbitrary shape in threedimensional space. 
For the sake of simplicity the movement of both particles is modeled as relative motion.
This means one of the particle is fixed in the center of the coordinate system (named particle A), while the other particle moves relatively from the perspective of the fixed particle (named particle B).
The space around the fixed particle is divided by a spherical surface of radius b into an outer region $r > b$ and an inner region $r < b$. \\
In order to estimate the rate of the diffusional association \textit{k}, one usually simulates several thousands of BD trajectories and measures the fraction of ``reactive'' trajectories. 
The rate of association is then computed by the product of the analytically computed rate k(b) and the probability $\beta^\infty$ estimated from simulations: 
\begin{align}
 k = k_D (b) \beta^\infty .
\end{align}
The rate $k_D(b)$ is the steady state rate at which particles with distance $r>b$ collide with the spherical surface at $r=b$. It is defined as  
\begin{align}
k_D(b) = 4\pi \left[ \int_b^\infty \frac{\exp(E(r)/k_BT)}{r^2D} \text{dr} \right]^{-1} \text{,}
\end{align}
where $E(r)$ is the centrosymmetric interaction energy between the particles depending on their separation distance r.
If the particles are noninteracting spheres, i.e. no forces or hydrodynamic interactions exist for $r>b$, $k_D(b)$ is given by the Smoluchowski result: 
\begin{align}
  k_D(b) = 4\pi D_0 b, 
\label{eq:smoluchowski}
\end{align}
with $D_0$ being the standard Smoluchowski mutual diffusion coefficient. 
The quantity $\beta^\infty$ on the other hand is the probability that particles reaching a separation $b$ will subsequently have at least one collision with the fixed particle rather than escape into infinity. 
$\beta^\infty$ is usually given by 
\begin{align*}
 \beta^\infty = \frac{\beta}{1-(1-\beta)\Omega}, \quad \text{with} \quad \Omega = \frac{k_D(b)}{k_D(q)}.
\end{align*} 
Although the development of the NAM method dates back to the 1980's, most of today's used methods are still derived from this simple method. 
However large improvements have been made in adequately modelling the interparticle forces and in optimizing, i.e. steering the simulation of the BD-trajectories. 

\paragraph{Modeling of Force Fields}

Molecules carry partial charges which generate an electrostatic potential at the molecule's surface. 
As soon as two molecules overcome a certain distance, their diffusional movement becomes biased by the forces that their electric charges exert on each other. \\
Such intermolecular forces strongly influence the speed of association. 
In diffusional association these forces are typically given by the sum of the electrostatic and exclusion forces.
Electrostatic interactions and hydrophobic desolvation typically speed up the association, whereas charge desolvation (\ie a penalty for bringing a charge close to a low dielectric molecular cavity) and steric repulsion terms lower the speed of association. 
As a result, the association rate of different pairs of molecules may range between $10^3 - 10^9 Ms^{-1}$. \\
Electrostatic interactions are typically computed by the Poisson-Boltzmann equation (PBE). 
There exist several methods exploiting numerical solvers tackling the PBE even for larger systems \cite{Davis:1989}. 
However, such numerical methods are still too time consuming to be applied in each step of the million steps of a BD trajectory. 
The search for a simplistic, but accurate approximation of intermolecular forces is therefore one of the most challenging problems in the modeling and simulation of diffusional association. \\
A variety of different electrostatic models have been used in BD simulations. 
They mainly vary in the level of detail employed in the modeling of the charge distribution and the approximation used to present the electrostatic interactions of the system. \\
One of the most basic method is the test charge method. 
Here the electrostatic potential is computed for the fixed protein (particle A) on a grid.
The second molecule (particle B) is considered as a collection of point charges in the solvent dielectric that moves on the potential grid of the fixed protein. 
The low dielectric and ion exclusion volume of the second molecule are ignored. 
Although the method is rather coarse, it has been the standard method for a long time. \\
More accurate electrostatic models, with almost the same computational effort, have been proposed by Gabdoulline and Wade \cite{Gabdoulline:1996} and by Beard and Schlick \cite{Beard:2001}. 
Gabdoulline and Wade extended the standard ``test-charge'' approach to the effective charge model. 
Here the electrostatic potential is first calculated by a numerical solution of the PBE and for each molecule separately.
Afterwards, the molecules are parameterized by fitting effective charges on the molecules such that the previously calculated external potential is fully represented. 
This has the advantage of getting a simple, but rather realistic electrostatic model of the molecules, by solving the PBE numerically only once.
Beard and Schlick have developed a similar approach to reproduce the electric field of the PBE. 
They parameterize the molecule by effective surface charges and apply this approximation in a Debye-H\"uckel model.\\
The introduction of a charge desolvation term brought further progress in the development of modeling techniques for DA. 
The charge desolvation term describes the penalty due to bringing a charge close to a low dielectric molecular cavity. 
A combination of the effective charges approach, the charge desolvation term and an appropriate reaction criteria has been used as a single protocol, in order to simulate the association of several protein pairs with the same model \cite{Gabdoulline:2001}.

\paragraph{Sampling}

Many molecular systems contain energy barriers, that have to be crossed before the particles can associate. 
Such energy barriers are e.g. one or several unfavorable states with respect to steric repulsion of electrostatic forces. 
In these cases the fraction of reactive trajectories is very small. 
Consequently one has to perform an infeasible amount of BD trajectories until a converged association rate is reached. 
\\
With the Weighted Brownian dynamics (WEB) \cite{Huber:1996}, Huber and Kim adapted the NAM method to overcome reaction barriers. 
Instead of modeling the association of a single pair of particles, the WEB method replaces the moving particle of the NAM method by an ensemble of weighted pseudo-particles and divides the configuration space into a number of bins along the reaction coordinate. \\
Each bin is assigned a probability based on a  Boltzmann distribution and initially filled with a number of weighted particles. 
During simulation the pseudoparticles are split and combined in order to equally sample the bins, i.e. all parts of the configuration space. 
WEB produces consistent results compared to the NAM approach. For fast associating processes WEB needs approximately the same computational time as the NAM method. However, for slow associating processes, WEB gains a 3-fold speed up in comparison to the NAM method \cite{Rojnuckarin:2000}.


\section{The structure of FADA}

In this section, we will introduce the Flexible Architecture for Diffusional Association (\fada) which has been implemented in Java.
Therefore, we will summarize the requirements for \fada, resulting from state of the art of Brownian dynamics modeling described in the previous section.
Based on that, we will explain the interplay of the architecture's basic components, while the following two sections will get into more detail about the components themselves.


\subsection{Requirements}

In order to create a flexible architecture that allows to implement a variety of DA methods, we need a basic model illuminating the biological setting of association processes for all possible implementations. 
Such a model needs to describe two or more molecules of some shape and certain properties in solution. 
These molecules move under constraints.
Thus, their motion has to be modeled by a certain movement function, which defines velocity and direction of the movements with respect to the given constraints. 
For molecular systems these constraints mainly refer to the potential of mean forces, which incorporates ionic strength of the water molecules as well as intrinsic interactions between the particles. \\
To summarize, the general model of DA consists of three main components: Particles Model (Individual), Movement Model, Potential of Mean Force Model. 
These components have to be implemented by every DA approach and thus form the basis of our DA model. \\
For each DA approach, these components are refined by specifying the mathematical model in terms of rules and parameters. 
A language should be available, to specify models in a user-friendly way.
A specific simulator will then be responsible for the execution. 
It should be possible to apply different simulator components that allow different execution strategies. 
This is of great use for model validation (since biased components can be identified \cite{Leye2010}) and it also provides the possibility of automatically choosing the best algorithm (simulator) for the given simulation model \cite{Ewald:2009b}. 
Consequently, a flexible experimentation workflow is required, that allows the configuration of model and simulator, as well as experiment design and analysis methods to exploit, extend and compare existing techniques.
%

\subsection{Basic components of \fada}

\begin{figure}[htb]
	\centering
		\includegraphics[width=10cm]{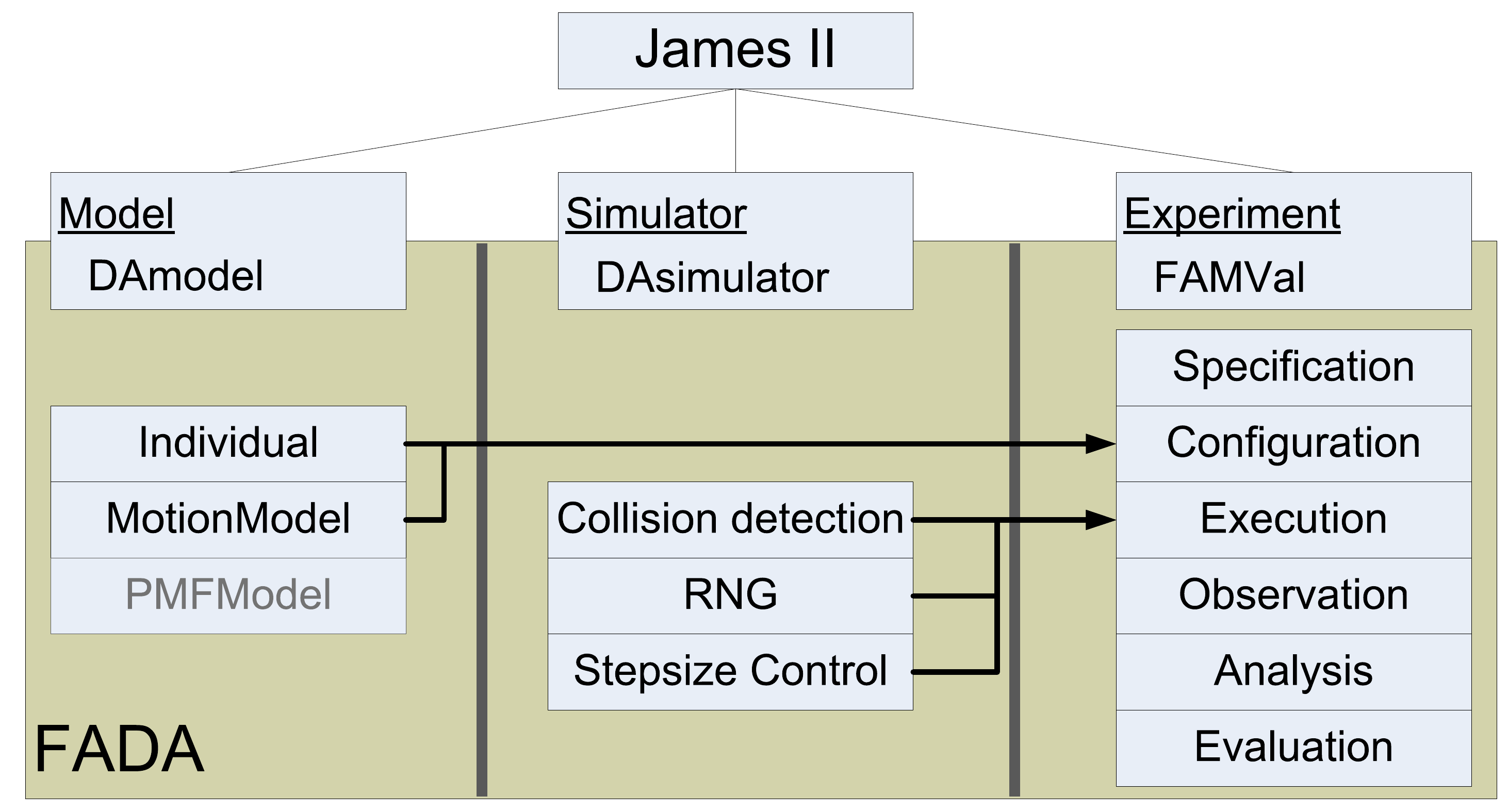}
\caption{The structure of \fada.}
\label{fig:FADA}
\end{figure}

The key point of \fada's design is the clear distinction between model, simulator and experiment.
It follows the philosophy of \james \cite{Himmelspach2007, Himmelspach2008}, whose plug-in structure is the foundation for \fada.
Hence, for each of the three steps different components and methods can be created, allowing a flexible combination and application of them (see Figure \ref{fig:FADA}).

However, since the area of application for \fada is the DA process, some of these components have to follow a specific structure.
A model can be specified with the SpacePi-Calculus (see Section \ref{modeling}).
It comprises the two basic components: Individual, to describe particles and MotionModel to describe their movement.
A PMFModel component, describing additional properties of the particles is under development (see Section \ref{outlook}). 
Accordingly, simulation engines executing those models can be created.
The simulator components may be exchanged and extended by exploiting the plug-in architecture.
So far a DA simulator comprises components for the generation of random numbers, for the collision detection, and for stepsize control if the collision detection works time-stepped.
Validation experiments can be configured and executed using the validation environment \famval \cite{Leye2010}.
Those experiments follow six basic steps (see Section \ref{famval}).
For each step, methods and techniques may be created, extended, or exchanged (again by exploiting the plug-in architecture) in order to allow a flexible configuration of experiments.

\section{Modeling and simulation with FADA}
\label{modeling}

In the following we will specify the NAM method in SpacePi \cite{John2008} and thereby show the exemplary definition of a simulation model and illustrate the functioning of FADA. \\
SpacePi is an extension of the $\pi$-Calculus \cite{Milner:1999}, that incorporates time and space. 
The $\pi$-Calculus allows the definition of multiple parallel processes and their interactions.
However, in contrast to standard $\pi$, SpacePi allows to assign a position (in two- or three-dimensional space) to a process. 
Further a movement function can be assigned to each process, describing its spatial motion.
Constraints can be defined, such that the possible interactions of the processes depends on the spatial configuration of the system, \ie the position, or precisely the relative distance of the two particles. 
SpacePi thus provides all important features, that we need to specify the NAM model 

\subsection{Model Specification for Association Processes}
\label{sec:model}


Before the NAMmodel can be specified in SpacePi, it is necessary to introduce the basic notations of the SpacePi formalism. 
SpacePi extends the $\pi$ process with spatial notion to allow free movement of the processes. 
Each process P is associated with certain position $\overrightarrow{p} \in V, V \in R^d$, where $\overrightarrow{p}$ is a vector space with norm $\|\cdot \|_v$. 
The modeling of movements demands the inclusion of time, as the speed of the motion could not be expressed otherwise. 
In SpacePi time is expressed by intervals ($\delta_t$) and the communication takes place during the time intervals. 
Therefor a single rule is introduced to ensure the progress from time interval to time interval, while all other rules define the activities taking place during each interval. 
During a time interval the velocity and direction of motion are constant.
Thus the movement function reads $m~ = ~ V \times X \rightarrow V$, which takes the current position of the process and a set of additional parameters X and calculates the new position with respect to X. 
Note that X as well as a function for selecting parameters $\chi \in X$ needs to be provided by the model.
The movement function generates a target vector of the process.
The target vector is added to the current position in order to receive the target position $\overrightarrow{t} = \overrightarrow{p} + m(\overrightarrow{p}, \chi)$, which is the position reached after the time interval $\delta_t$. \\
Based on the previous definitions a process can be associated with its current position and a movement function m.
We thus write a process in SpacePi as $P_m^{\overrightarrow p}$. 
The formal description of a SpacePi process is as follows: 
\begin{align*}
 P_m^{\overrightarrow p}~::=~\sum_{i\in I}~ \pi_i~ .~ P_{i,~m_i}^{\overrightarrow p_i}~\mid~P_{1,~m_1}^{\overrightarrow p_1}~ \mid ~P_{2,~ m_2}^{\overrightarrow p_i} ~ \mid ~ (new~x) P_{m}^{\overrightarrow p} ~ \mid ~ nil
\end{align*}
Nil is a shorthand for $nil_{m(\overrightarrow 0, \chi)}^{\overrightarrow 0}$ and describes an empty process without movement and a position at the point of origin of V. \\
The action $\pi$ refers to a channel through which the concurrent processes may communicate. 
Sending and receiving x over the channel ch with radius r is denoted by $ch !(x,r)$ and $ch ? (x,r)$, respectively. 
Two processes may only communicate, if their distance is closer than the given action radius $r$. 
A communication action with $r=\infty$ resembles the semantics of the $\pi$ calculus, i.e. the channel has no spatial restriction.
Based on the given notations of the SpacePi formalism, we can now specify the NAM model.
Our system consists of two Brownian particles moving in $\mathbb R^3$.  
The motion is described in terms of Brownian motion. 
We thus define the movement function as $mov(P_i, pos, t) = pos^t$ with $pos^t(P_i) = pos^{t-1}(P_i) + S$, and S being a $\mathbb{R}^3$ random variables drawn from a normal distribution with mean $\langle S \rangle = 0$ and derivation of $\langle S^2 \rangle = 2D\delta_t$. 
D is the diffusion coefficient, which has to be provided by the parametrization of the model. 
The movement function resembles the mean displacement of a Brownian particle based on the Ermak-McCammon Equation \cite{Ermak:1978} under no influence of interparticle forces. 
There are two events, that we want to monitor. 
Either the two particles react, i.e. collide, or the particles diffuse into infinite seperation.\\
The specification of the model in SpacePi is shown in figure \ref{fig:example}. 
Please note that no potential mean force model is specified. 
The model consists of three different processes $FixedParticle$, $MovingParticle$ and $ExitParticle$.
The first two processes describe the two particles under consideration. 
Since we model the movement of both particles as relative motion, only the $MovingParticle$ is associated with a (Brownian) movement function ($bMove$). 
The process $FixedParticle$ is fixed in the center of the coordinate system and does not move. 
The expression $coll?(\sim, r) $ describes the collision event, which occurs as soon as the two particles reach a distance smaller than the given action radius r. 
Thereby the channel's name, the radius of the complementary channel actions, and the location of the processes are taken into account.
After the collision, i.e. reaction event, the processes terminate. The third process $ExitParticle$ is an auxillary process that allows to realize the escape event. 

\begin{figure}[h!]
\small
\ovalbox{
	\cornersize{0}	
	\begin{minipage}{\textwidth}
	\vspace{4pt}
	\textbf{Position declarations}\\
	$pos_F ~:=~ x = 0 ~ \wedge ~ y = 0 ~ \wedge ~ z = 0$ \\
	$pos_M ~:=~ rand(x,y,z) ~ s.t. ~ (x^2 + y^2 + z^2) = b$ \\
	$pos_E ~:=~ rand(x,y,z) ~ s.t. ~ (x^2 + y^2 + z^2) = q$ \\
	\newline
	\textbf{Radius declarations}\\
	$r_{react}$ = 10 \\
	\newline
	\textbf{Potential of mean force declarations} \\ 
	$f_{pmf}: \text{not defined}$ \\  
	\newline
	\textbf{Motion declarations}\\
	bMove(): $ \dot{x}^2 + \dot{y}^2 + \dot{z}^2 < q, \quad x = pos_E(x) ~ \wedge ~ y = pos_E(y) ~ \wedge ~ z = pos_E(z)\ \text{otherwise} $\\
	\newline
	\textbf{Process definitions} \\
	$ FixedParticle = coll? (\sim, r).0 $  \\
 	$ MovingParticle[bMove] = coll! (\sim, r).0$ \\
	$ ExitParticle =  coll? (\sim, r).0 $\\
	\newline
	\textbf{Initial process} \\
	$ FixedParticle  ~ | ~ MovingParticle ~ | ~ ExitParticle$ 
	\newline
	\end{minipage}
}	
\caption{\ A simplistic SpacePi model based on the NAM model for diffusional association of two particles. An expression $ch?(\sim, r)$ denotes that an empty message is to be received on channel ch with radius r.}
\label{fig:example}
\end{figure}
In order to model the escape of the moving particle, the NAM model defines a separation distance q at which the two particles have diffused into infinite separation. 
As neither $\pi$ nor SpacePi calculus offer a way to represent boundaries, we incorporate the distance constraint into the movement function. 
The idea is to place the $ExitParticle$ outside of the q-surface and change the movement function of the $MovingParticle$, such that $MovingParticle$ will collide with $ExitParticle$ as soon as it reaches the separation distance q. 
We thus extend the movement of the process $MovingParticle$ as follows $mov(P_M, pos, \delta_t) = pos^t$ with $pos^t(P_M) = pos^{t-1}(P_M) + S, \ (\text{iff} ~ \|pos^t(P_M) + S\| ~ < ~ q, ~ pos^t(P_M) = pos(P_E)$ otherwise$)$. 
Thereby $P_M$ stands for $MovingParticle$ and $P_E$ for $ExitParticle$. 
Due to the direct displacement of $MovingParticle$ to the position of $ExitParticle$, the two particles collide instantaneously when $MovingParticle$ has passed the distance q. 
The collision of both processes $coll?(\sim, r)$ is similar to the collision action of $FixedParticle$ and $MovingParticle$. After this collision, i.e. escape event, the processes also terminate. 
Although the specified model is far from representing a realistic biological association process (for this a potential mean force model has to be included), the model serves as a good example to present the general idea, of how association processes can be expressed by a model formalism, evaluated by a simulator and finally be validated through validation experiments. 
Further the simplicity of the model allows us to calculate an accurate analytical solution, to validate our approach. 

\subsection{Simulation of the model}

The simulator evaluates a parameterized instance of the described simulation model. 
For the simulation of the trajectory we use a simple step simulator, which calculates the mean displacement of the moving particle in each time step and checks for collision events. 
Therefore the simulator needs to implement a Random Number Generator (RNG) and a collision detection algorithm, which are realized as plug-ins. 
A separate implementation of these components (RNG and collision detection) allows to use different simulator configurations by applying different algorithms in each component. 
So far, \james offers over 10 different RNG implementations, that may all be used by \fada. 
For the collision detections, we implemented two approaches, a time stepped and an event-triggered one.
The time-stepped one, measures the distance between the particles at distinct time steps and reports a collision when the distance is $0$.
To minimize the amount of undetected collisions (\ie two particles have a distance $= 0$ inbetween two time steps but $> 0$ at the steps) , a stepsize control is required for which we implemented two versions as well.
The first one creates a constant stepsize over the whole simulation run. 
The second version determines the stepsize depending on the distance of the moving particle to the reaction or escaping sphere (the lesser the distance the smaller the time step).
The event-triggered collision detection precalculates the time point at which the particles should collide, depending on the actual movement and position.
If no movement update happens before the calculated time point, the reaction happens. 
This variant requires more complicated calculations (compared to the time-stepped version) but produces an exact simulation. 
Another benefit of the given separation of model and simulation, is the chance to optimize each component separately (\eg by advanced step-size adaptations, heuristics, parallelization, etc.).
This allows the integration of specialized solutions created by experts.
Furthermore, additional features of the model can be reflected by new components for the simulator (\eg a component handling the PMF model). 

\section{Experimentation with FADA}
\label{famval}
To conduct experiments with the BD model and simulator, \fada exploits the validation environment \famval.
\famval is based on six steps that have been identified as being crucial for structuring validation experiments \cite{Leye2009, Leye2010}.
In this section, we describe those steps and explain how each of them has been adapted in order to test the reaction rate of the BD model from Section \ref{sec:model} against the results of the state of the art and present some results. 

\subsection{Important steps of a validation experiment}

The first and most important step is the specification of requirements for the model to be validated. 
Thereby, the parameters for the whole experiment are defined, influencing each of the following steps.

The second step is the configuration of the model, where interesting points in the model's parameter space are selected in order to achieve the required information about the validity of the model. 

The third step is the execution of the model.
Valid simulation engines and components are required for a reliable simulation result.
Therefore, simulation runs have to be repeated with different simulator settings, to identify those biasing the simulation output due to approximation issues, bugs or side effects.

The fourth step is the observation of the model execution, to produce the required simulation output.
In the ideal case, only those properties of the model are observed, that are required to make a decision about whether the model fulfills the requirements or not.
More observations would produce overhead, less would hamper the analysis of the results.

The analysis of the observed data is the fifth step.
It comprises two parts: the analysis of a single simulation run and the analysis of a set of replications. 
In general, the second part is based on the results of the first one, but occasions exist where just one of them is necessary (\eg the comparison of trajectories over a set of replications to calculate the monte-carlo variability).
To execute a proper analysis, the required amount of simulation end times and required replications has to be determined.

Finally, the last step is the evaluation. 
It takes the result of the analysis as well as the requirements into account. 
Feedback for the configuration step may be produced, leading to iterative experiments which can be exploited for parameter estimation and optimization tasks in general.

\subsection{An example experiment}

The comparison of the simulation results of a newly developed model to those of a well-established model is a typical example for a model validation experiment \cite{Sargent2008}.
We describe such an experiment to validate the model of section \ref{sec:model} following the described structure of \fada. \\ 
The basic demand for the model is that its estimated rate constant $r_{s} = k_D(b) \beta^\infty$ matches the analytical solution $r_{a} = 4\pi Da$, where $a$ is the reaction criterion, \ie the minimum separation required for the two particles to react. 
In this case, the probability $\beta$ that is required to estimate the rate $r_{s}$ can be directly computed by: 
\begin{align*}
 \beta_{a} = \frac{a}{b}\cdot\frac{q-b}{q-a}
\end{align*}
We can thus use the probability rate $\beta_{s}$ as requirement, instead of the reaction rate constants. 
Hence, for a valid model the rate $\beta_{s}$ of simulation runs terminating with a reaction of the two particles, needs to match the rate $\beta_{a}$. 
For our setting, the analytical rate $\beta_{a}$ has the value of 0.111. 
Besides the given rate, an error tolerance $e$ is required, denoting the maximum distance between $\beta_{s}$ and $\beta_{a}$ for the model to be considered valid, with a given confidence $c$.
In our experiment we set $e$ to 0.05 and $c$ to 0.99.
\begin{table}[hb!]
\small
\centering
\begin{tabular}{|c|c|}
\hline
\textbf{Validation steps} & \textbf{Realization} \\ 
\hline
Specification &  determine $\beta_{a}$, $e$, $c$\\
 & configure the following steps\\
\hline
 & diffusion coefficient D\\
Configuration  & initial particle distance b\\
	      & minimal collision distance a\\
	      & minimal exit distance q\\
\hline
          & Mersenne Twister/Default RNG \\ 
Execution & time stepped/event triggered \\ 
          & fixed/adaptive steps \\ 
\hline
Observation & end state \\
\hline
Analysis & calculate $\beta_{s}$ by taking the average over all runs \\
 & calculate the required replications \\
\hline
Evaluation & compare  $\beta_{s}$ to $\beta_{a}$ \\
\hline
\end{tabular}
	\caption{Overview over the steps in a validation experiment and the realization to validate the BD model.}
	\label{fig:SimpleResults}
\end{table}
Since the reaction rate constant is only dependent on the reaction criterion, the parameter b, q, and D may be chosen arbitrarily.  
Typically q is chosen sufficiently larger than b, in order to allow a significant fraction of trajectories to react. 
For the configuration we used a fixed minimum reaction criterion a = 10{\AA} and fixed distances b = 50{\AA}, and q = 100{\AA}.
The radius of the particles has been set to 4{\AA}. 
This choice of parameters is based on the experiment settings used by \cite{Rojnuckarin:2000}.
Since we selected the parameters by hand (in contrast to using an algorithm, like a parameter sweep), they have to be reflected in the definition of the requirements. \\
For the execution of the model we used two different RNG's, the standard Java RNG and the Mersenne Twister, to investigate, whether a change of the RNG may have an influence on the simulation results.
Furthermore, for executing the model we tried both implemented collision detection variants and for the time-stepped one both stepsize controls.
The default stepsize was 0.1 milliseconds and the adapted one was 0.01 milliseconds.
To calculate the association rate of two molecules just one observation is of importance, which is the fact whether the two particles have reacted or not.
Therefore, the end state of reach simulation run has to be collected.
The end of a simulation run is reached, either if the two particles collided or if the distance between them reached a given threshold.
No additional effort had to be put in the analysis of the single simulation runs.
The information about the end state of a run does not have to be post-processed in order to calculate $\beta_{s}$. 
This value is created during the analysis of the configuration (\ie the replications with the same model parameter configuration).
Thereby, a list holding the end state of each simulation is maintained and after the required replications have been finished, the number of runs where the particles have reacted is summed up and divided by the count of all replications.
It is not sufficient to skip the list and calculate the sum on the way, since the intermediate results are used to calculate the standard error, which is required to calculate the count of necessray replications to ensure that the resulting rate lies in the given error range $e$ with confidence $c$ \cite{Chung2004}. \\
The evaluation basically comprises the comparison of the given rate to that, produced by the simulation.
If the difference lies in the given tolerance $e$, the model can be considered valid for the used parameter configuration.
Since the configurations are selected by hand at the beginning of the experiment, no loop and therefore no feedback from the evaluation to the configuration is required.

\begin{figure}[htb]
	\centering
		\includegraphics[width=10cm]{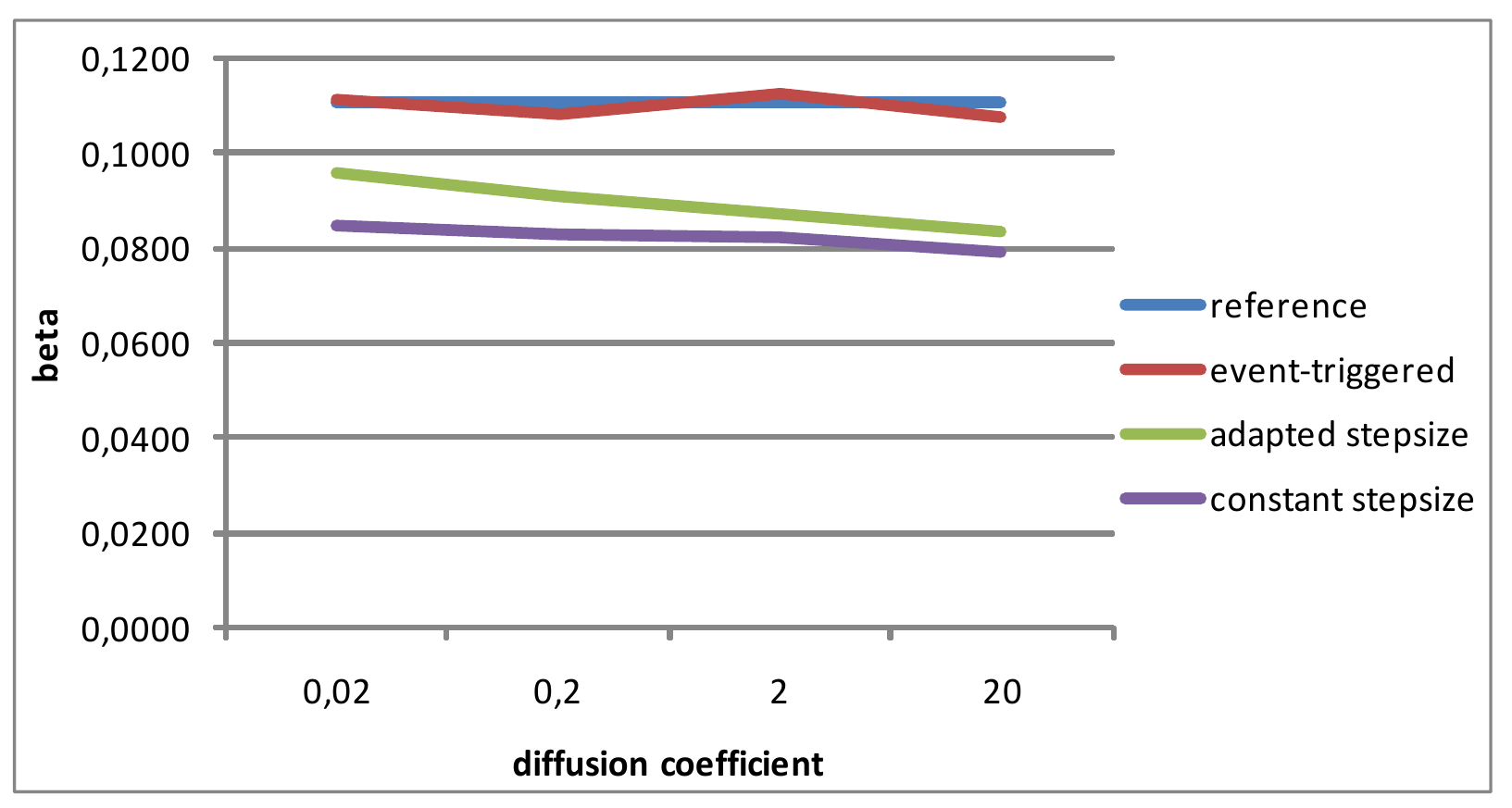}
  \caption{Results of the validation experiment.}
\label{fig:results}
\end{figure}

\paragraph{Experiment Results}
%
Figure \ref{fig:results} shows the results of the described experiment.
The exchange of the random number generator did not make any significant difference for the results.
Therefore, we just present the results of the Mersenne Twister. 
The simulations with the event-triggered collision detection produced association rates, that did lie inside the error tolerance.
However, a significant difference exists between the results produced by the time-stepped collision detection and the analytical result.
As the diffusion coefficient is increased, the association rate produced by the simulation is decreased, while the reference rate stays constant.
The differences are higher with the constant stepsize than with the adaptive stepsize.
They can be explained by the less precise simulation with the time-stepped collision detection.
If the collision happens between two steps and the two particles do not overlap in the second step, the collision can not be detected.
As the particles have a lower diffusion coefficient and therefore move slower \ie they pass a smaller distance between two time steps, the probability of missing a collision is lower.
Similarly, if the time step is adapted when a particle approaches the other one (and therefore again the passed distance between two time steps is decreased), less collisions are missed. \\
The experiment shows the importance of testing different simulation engine configurations during a validation experiment.
The type of the collision detection as well as the stepsize control had a high impact on the experiment results.
If for instance just the time-stepped detection with the constant stepsize would have been used, the biased results would have led to the conclusion that the given model configuration is invalid which it was not.

\section{Outlook}
\label{outlook}

The model presented and used here for testing purposes is simplistic. 
In order to include more realistic dynamics, \eg for interparticle forces and reaction criteria, it is necessary to extend the model described in section \ref{sec:model}. 
Since SpacePi allows an arbitrary definition of the movement function, different approaches for the modeling of the potential mean force can be integrated. 
Therefore, one has to define the corresponding functions, \ie declaring the potential of mean force model (see figure \ref{fig:example}), that calculate the impact of the interparticle forces, while taking in account the position of the two particles. 
The value of this force field evaluation is then used in order to parametrize the movement function of the moving particle.
However, in addition it is necessary to parameterize the processes in terms of (local) charges of the particles and also introduce a parameter for the ionic strength of the surrounding solvent. \\
Another interesting, but challenging step is the integration of more complex reaction criteria. 
Therefore a more detailed parametrization of the processes in terms of shape and size is necessary. 
A promising approach, to solve this problem has been reported by \cite{Schaefer:2009}, who introduced labels to the SpacePi formalism. 
%
In our validation experiment we estimated the association rate of the presented model. 
It could be easily validated as the value is analytically known. 
However, in any realistic application of this procedure this will not be the case. 
Thus, the validity of the used model has to be ensured.
Therefore, it is necessary to gain confidence in the model with adequate methods.
One way to achieve this confidence is to check the plausibility of the model's behavior.
In the following we want to give an overview over methods, that might be applied in order to check the plausibility of the BD model. \\
During the specification step the requirements the model has to fulfill, have to be defined.
Those requirements have to be pointed out with respect to the overall goal of the model - the calculation of association rates.
Therefore it is important to identify those properties of the model, that influence the association rates and that can be checked for their plausibility.
For instance, the end states of the simulation runs do not give good insights into the validity of the model since a reference association rate to check the plausibility of their occurrence is not available.
The plausibility of a model does not just depend on the behavior of one single configuration.
If neighbor configurations  show unexpected behavior, problems in the model structure might exist that could influence the results of the configuration under investigation.
Hence, experiments with different configurations can give new insights about the model structure.
Furthermore, the reusability of the model is increased, if different parameter settings have been checked for their plausibility.
If new particles with different properties have to be investigated it might be sufficient to adapt the model parameters instead of creating an entire new model. 
However, those parameters have to be covered during the validation process.
An algorithm exploring the parameter space \cite{Calvez2005} is required therefore. 
Future parameters of the model (\eg forcefield, rotational diffusion, ionization of solvent) have to be considered in this phase. \\
For the execution of the model, different simulation engines and components have to be provided, to investigate bias produced by different algorithms. 
To get decent information about the behaviour of the modeled particles, more complex observations are required.
As mentioned before, the end state of the simulation is not sufficient anymore.
However, more detailed information about the behaviour particles, like \eg their movement trajectory could give hints about whether issues exist, that could bias the potential reactions.\\
To analyze the simulation output results, techniques to analyze simulation traces \cite{Kleijnen2000} may be used.
Furthermore, techniques from Runtime Verification \cite{Bensalem2009} and simulation-based model-checking \cite{Fages2007} could be well suited.
Thereby, expected paths of the particles would be coded into LTL-formulae and the model-checker could check, whether the movement trajectories produced by the simulation runs follow the path.
This analysis technique could be used twofold for validation experiments.
On the one hand, it could by used directly to validate the reaction events, \eg something must be wrong if the trajectories of the particles should have led to a reaction, but now reaction occurred.
On the other hand, it could be used, to validate certain parts of the model, \eg something must be wrong if the trajectories of the particles contradict their movement functions. \\
The evaluation of the analysis results is not constrained.
A simple comparison of the results, to given specifications is imaginable, which however would not offer hints about the problem if the comparison turns out to be negative.
One technique giving such hunts and supporting model debugging, is face validation.
The analysis results are post-processed to create figures that support the modeler during the process of finding errors in the model.
Those figures might facilitate the browsing through model structure and experiment results \cite{Unger2009} or display interactions between the model's components \cite{Kemper2006}.

\section{Conclusion}

We introduced the basic concept of \fada, an architecture for the modeling, simulation, and experimentation of diffusional association processes.
\fada uses SpacePi-Calculus expressions for the definition of models.
Those models comprise two particles, whose reaction behaviour is under investigation.
So far, this behaviour is solely based on the movement functions of the particles, but additional features (\eg interparticle forces, detailed molecule representation, decent reaction criteria) are under development.
We proposed a simple simulation engine relying on the three components collision detection, random number generator and stepsize control.
According to additional model features, the simulator will be adapted by new components in the future.
For the experimentation, \fada distinguishes six steps in order to conduct a validation experiment.
Exemplarily, we described the realization of those steps for the validation of the association rate of a BD model.
The results of this experiment showed that the simulation engine has an impact on the behaviour of the simulated model and that different simulator configurations have to be used in order to sort out invalid ones.
With new features of the model, the realized steps of the experimentation process have to be adapted.
This is especially important, if the model has to be validated with no reference association rate at hand, which is the case, if the model should be used to estimate association rates.
For both tasks, the validation of DA models as well as the estimation of reaction rates based on them, \fada can be used.

\bibliographystyle{eptcs} 
\bibliography{Literatur.bib}

\begin{thebibliography}{10}
\providecommand{\bibitemstart}[1]{\bibitem{#1}}
\providecommand{\bibitemend}{}
\providecommand{\bibliographystart}{}
\providecommand{\bibliographyend}{}
\providecommand{\url}[1]{\texttt{#1}}
\providecommand{\urlprefix}{Available at }
\providecommand{\bibinfo}[2]{#2}
\bibliographystart

\bibitemstart{Beard:2001}
\bibinfo{author}{DA~Beard} \& \bibinfo{author}{T.~Schlick}
  (\bibinfo{year}{2001}): \emph{\bibinfo{title}{Modeling salt-mediated
  electrostatics of macromolecules: the discrete surface charge optimization
  algorithm and its application to the nucleosome}}.
\newblock {\sl \bibinfo{journal}{Biopolymers}} \bibinfo{volume}{58}, pp.
  \bibinfo{pages}{106--115}.
\bibitemend

\bibitemstart{Bensalem2009}
\bibinfo{editor}{Saddek Bensalem} \& \bibinfo{editor}{Doron~A. Peled}, editors
  (\bibinfo{year}{2009}): \emph{\bibinfo{title}{Runtime Verification}}.
\newblock \bibinfo{publisher}{Springer}.
\bibitemend

\bibitemstart{Berg:1985}
\bibinfo{author}{O.~G. Berg} \& \bibinfo{author}{P.~H. von Hippel}
  (\bibinfo{year}{1985}): \emph{\bibinfo{title}{Diffusion-Controlled
  Macromolecular Interactions}}.
\newblock {\sl \bibinfo{journal}{Annual Review of Biophysics and Biophysical
  Chemistry}} \bibinfo{volume}{14}(\bibinfo{number}{1}), pp.
  \bibinfo{pages}{131--158}.
\bibitemend

\bibitemstart{Berry:2002}
\bibinfo{author}{H.~Berry} (\bibinfo{year}{2002}): \emph{\bibinfo{title}{Monte
  Carlo simulations of enzyme reactions in two dimensions: fractal kinetics and
  spatial segregation}}.
\newblock {\sl \bibinfo{journal}{Biophys. J.}} \bibinfo{volume}{83}, pp.
  \bibinfo{pages}{1891--1901}.
\bibitemend

\bibitemstart{Briggs:1995}
\bibinfo{author}{J.~Briggs}, \bibinfo{author}{J.~Madura},
  \bibinfo{author}{M.~Davis}, \bibinfo{author}{M.~Gilson},
  \bibinfo{author}{J.~Antosiewicz}, \bibinfo{author}{B.~Luty},
  \bibinfo{author}{R.~Wade}, \bibinfo{author}{B.~Bagheri},
  \bibinfo{author}{A.~Ilin}, \bibinfo{author}{R.~Tan} \& \bibinfo{author}{J.~A.
  McCammon.} (\bibinfo{year}{1995}): \emph{\bibinfo{title}{University of
  Houston Brownian Dynamics Program User's Guide and Programmer's Manual
  Release 5.1.}}
\newblock \bibinfo{publisher}{Cambridge University Press, New York.}
\bibitemend

\bibitemstart{Calvez2005}
\bibinfo{author}{B.~Calvez} \& \bibinfo{author}{G.~Hutzler}
  (\bibinfo{year}{2005}): \emph{\bibinfo{title}{Parameter Space Exploration of
  Agent-Based Models}}.
\newblock In: {\sl \bibinfo{booktitle}{Knowledge-Based Intelligent Information
  and Engineering Systems}}. pp. \bibinfo{pages}{633--639}.
\bibitemend

\bibitemstart{Chan:1984}
\bibinfo{author}{D.~Y. Chan} \& \bibinfo{author}{B.~Halle}
  (\bibinfo{year}{1984}): \emph{\bibinfo{title}{The
  Smoluchowski-Poisson-Boltzmann description of ion diffusion at charged
  interfaces}}.
\newblock {\sl \bibinfo{journal}{Biophys. J.}} \bibinfo{volume}{46}, pp.
  \bibinfo{pages}{387--407}.
\bibitemend

\bibitemstart{Chung2004}
\bibinfo{author}{C.~A. Chung} (\bibinfo{year}{2004}):
  \emph{\bibinfo{title}{Simulation modeling handbook: a practical approach}}.
\newblock \bibinfo{publisher}{CRC Press}.
\bibitemend

\bibitemstart{Davis:1989}
\bibinfo{author}{M.~E. Davis} \& \bibinfo{author}{J.~A. McCammon}
  (\bibinfo{year}{1989}): \emph{\bibinfo{title}{Calculating electrostatic
  forces from grid-calculated potentials}}.
\newblock {\sl \bibinfo{journal}{J. Comp. Chem.}} \bibinfo{volume}{10}, pp.
  \bibinfo{pages}{386--391}.
\bibitemend

\bibitemstart{Ermak:1978}
\bibinfo{author}{D.L. Ermak} \& \bibinfo{author}{J.A. McCammon}
  (\bibinfo{year}{1978}): \emph{\bibinfo{title}{Brownian dynamics with
  hydrodynamics interactions}}.
\newblock {\sl \bibinfo{journal}{J. Chem. Phys}} \bibinfo{volume}{69}, pp.
  \bibinfo{pages}{1352--1360}.
\bibitemend

\bibitemstart{Ewald:2009b}
\bibinfo{author}{R.~Ewald}, \bibinfo{author}{S.~Leye} \& \bibinfo{author}{A.~M.
  Uhrmacher} (\bibinfo{year}{2009}): \emph{\bibinfo{title}{An Efficient and
  Adaptive Mechanism for Parallel Simulation Replication}}.
\newblock In: {\sl \bibinfo{booktitle}{PADS '09: Proceedings of the 2009
  ACM/IEEE/SCS 23rd Workshop on Principles of Advanced and Distributed
  Simulation}}. pp. \bibinfo{pages}{104--113}.
\bibitemend

\bibitemstart{Fages2007}
\bibinfo{author}{F.~Fages} \& \bibinfo{author}{A.~Rizk} (\bibinfo{year}{2007}):
  \emph{\bibinfo{title}{On the Analysis of Numerical Data Time Series in
  Temporal Logic}}.
\newblock In: \bibinfo{editor}{M.~Calder} \& \bibinfo{editor}{S.~Gilmore},
  editors: {\sl \bibinfo{booktitle}{CMSB}}. pp. \bibinfo{pages}{48--63}.
\bibitemend

\bibitemstart{Gabdoulline:1996}
\bibinfo{author}{R.~R. Gabdoulline} \& \bibinfo{author}{R~Wade}
  (\bibinfo{year}{1996}): \emph{\bibinfo{title}{Effective charges for
  macromolecules in solvent}}.
\newblock {\sl \bibinfo{journal}{J Phys Chem}} \bibinfo{volume}{100}, pp.
  \bibinfo{pages}{3868--3878.}
\bibitemend

\bibitemstart{Gabdoulline:1997}
\bibinfo{author}{R.~R. Gabdoulline} \& \bibinfo{author}{R.~C. Wade.}
  (\bibinfo{year}{1997}): \emph{\bibinfo{title}{Simulation of the diffusional
  association of barnase and barstar}}.
\newblock {\sl \bibinfo{journal}{Biophys. J.}} \bibinfo{volume}{72}, pp.
  \bibinfo{pages}{1917--1929}.
\bibitemend

\bibitemstart{Gabdoulline:2001}
\bibinfo{author}{R.~R. Gabdoulline} \& \bibinfo{author}{R.~C. Wade}
  (\bibinfo{year}{2001}): \emph{\bibinfo{title}{Protein-protein association:
  investigation of factors influencing association rates by Brownian dynamics
  simulations}}.
\newblock {\sl \bibinfo{journal}{J. Mol. Biol.}}
  \bibinfo{volume}{306}(\bibinfo{number}{5}), pp. \bibinfo{pages}{1139--1155}.
\bibitemend

\bibitemstart{Himmelspach2008}
\bibinfo{author}{J.~Himmelspach}, \bibinfo{author}{R.~Ewald} \&
  \bibinfo{author}{A.~M. Uhrmacher} (\bibinfo{year}{2008}):
  \emph{\bibinfo{title}{A flexible and scalable experimentation layer for
  {JAMES II}}}.
\newblock In: \bibinfo{editor}{S.J. Mason}, \bibinfo{editor}{R.R. Hill},
  \bibinfo{editor}{L.~Moench} \& \bibinfo{editor}{O.~Rose}, editors: {\sl
  \bibinfo{booktitle}{WSC}}. pp. \bibinfo{pages}{827--835}.
\bibitemend

\bibitemstart{Himmelspach2007}
\bibinfo{author}{J.~Himmelspach} \& \bibinfo{author}{A.~M. Uhrmacher}
  (\bibinfo{year}{2007}): \emph{\bibinfo{title}{Plug'n simulate}}.
\newblock In: {\sl \bibinfo{booktitle}{Spring Simulation Multiconference}}. pp.
  \bibinfo{pages}{137--143}.
\bibitemend

\bibitemstart{Huber:1996}
\bibinfo{author}{G.~A. Huber} \& \bibinfo{author}{S.~Kim}
  (\bibinfo{year}{1996}): \emph{\bibinfo{title}{Weighted-ensemble Brownian
  dynamics simulations for protein association reactions}}.
\newblock {\sl \bibinfo{journal}{Biophys. J.}} \bibinfo{volume}{70}, pp.
  \bibinfo{pages}{97--110}.
\bibitemend

\bibitemstart{Jeschke2008}
\bibinfo{author}{M.~Jeschke} \& \bibinfo{author}{R.~Ewald}
  (\bibinfo{year}{2008}): \emph{\bibinfo{title}{Large-Scale Design Space
  Exploration of SSA}}.
\newblock In: {\sl \bibinfo{booktitle}{CMSB '08: Proceedings of the 6th
  International Conference on Computational Methods in Systems Biology}}. pp.
  \bibinfo{pages}{211--230}.
\bibitemend

\bibitemstart{John2008}
\bibinfo{author}{M.~John}, \bibinfo{author}{R.~Ewald} \& \bibinfo{author}{A.~M.
  Uhrmacher} (\bibinfo{year}{2008}): \emph{\bibinfo{title}{A Spatial Extension
  to the Pi Calculus}}.
\newblock In: {\sl \bibinfo{booktitle}{Electronic Notes in Theoretical Computer
  Science}},  \bibinfo{volume}{194}. pp. \bibinfo{pages}{133--148}.
\bibitemend

\bibitemstart{Kemper2006}
\bibinfo{author}{P.~Kemper} \& \bibinfo{author}{C.~Tepper}
  (\bibinfo{year}{2006}): \emph{\bibinfo{title}{Traviando - Debugging
  Simulation Traces with Message Sequence Charts}}.
\newblock In: {\sl \bibinfo{booktitle}{QEST '06: Proceedings of the 3rd
  international conference on the Quantitative Evaluation of Systems}}. pp.
  \bibinfo{pages}{135--136}.
\bibitemend

\bibitemstart{Kleijnen2000}
\bibinfo{author}{Jack P.~C. Kleijnen}, \bibinfo{author}{Russel C.~H. Cheng} \&
  \bibinfo{author}{Bert Bettonvil} (\bibinfo{year}{2000}):
  \emph{\bibinfo{title}{Validation of trace-driven simulation models: more on
  bootstrap tests}}.
\newblock In: \bibinfo{editor}{J.~A. Joines}, \bibinfo{editor}{R.~R. Barton},
  \bibinfo{editor}{K.~Kang} \& \bibinfo{editor}{P.~A. Fishwick}, editors: {\sl
  \bibinfo{booktitle}{WSC}}. pp. \bibinfo{pages}{882--892}.
\bibitemend

\bibitemstart{Leye2009}
\bibinfo{author}{S.~Leye}, \bibinfo{author}{J.~Himmelspach} \&
  \bibinfo{author}{A.~M. Uhrmacher} (\bibinfo{year}{2009}):
  \emph{\bibinfo{title}{A discussion on experimental model validation}}.
\newblock In: \bibinfo{editor}{Adam Brentnall Ajith Abraham Richard~Zobel David
  Al-Dabass, Alessandra~Orsoni}, editor: {\sl \bibinfo{booktitle}{Proceedings
  of the 11th International Conference on Computer Modeling and Simulation}}.
  pp. \bibinfo{pages}{161--167}.
\bibitemend

\bibitemstart{Leye2010}
\bibinfo{author}{S.~Leye} \& \bibinfo{author}{A.~M. Uhrmacher}
  (\bibinfo{year}{2010}): \emph{\bibinfo{title}{A flexible and extensible
  architecture for experimental model validation}}.
\newblock In: {\sl \bibinfo{booktitle}{Simutools}}.
\bibitemend

\bibitemstart{Milner:1999}
\bibinfo{author}{R.~Milner} (\bibinfo{year}{1999}):
  \emph{\bibinfo{title}{Communicating and Mobile Systems: the Pi-Calculus}}.
\newblock \bibinfo{publisher}{Cambridge University Press}.
\bibitemend

\bibitemstart{Northrup:1984}
\bibinfo{author}{S.~H. Northrup}, \bibinfo{author}{S.~A. Allison} \&
  \bibinfo{author}{J.~A. McCammon} (\bibinfo{year}{1984}):
  \emph{\bibinfo{title}{Brownian dynamics simulation of diffusion-influenced
  bimolecular reactions}}.
\newblock {\sl \bibinfo{journal}{J. Chem. Phys.}} \bibinfo{volume}{80}, pp.
  \bibinfo{pages}{1517 -- 1524}.
\bibitemend

\bibitemstart{Rojnuckarin:2000}
\bibinfo{author}{A.~Rojnuckarin}, \bibinfo{author}{D.~R. Livesay} \&
  \bibinfo{author}{S.~Subramaniam} (\bibinfo{year}{2000}):
  \emph{\bibinfo{title}{Bimolecular reaction simulation using Weighted Ensemble
  Brownian dynamics and the University of Houston Brownian Dynamics program}}.
\newblock {\sl \bibinfo{journal}{Biophys J.}} \bibinfo{volume}{79(2)}, pp.
  \bibinfo{pages}{686--693.}
\bibitemend

\bibitemstart{Sargent2008}
\bibinfo{author}{R.~G. Sargent} (\bibinfo{year}{2008}):
  \emph{\bibinfo{title}{Verification and Validation of Simulation Models}}.
\newblock In: \bibinfo{editor}{S.J. Mason}, \bibinfo{editor}{R.R. Hill},
  \bibinfo{editor}{L.~Moench} \& \bibinfo{editor}{O.~Rose}, editors: {\sl
  \bibinfo{booktitle}{Proc. of the 39th WSC}}. pp. \bibinfo{pages}{157--169}.
\bibitemend

\bibitemstart{Schaefer:2009}
\bibinfo{author}{A.~Sch\"afer} \& \bibinfo{author}{M.~John}
  (\bibinfo{year}{2009}): \emph{\bibinfo{title}{Conceptional Modeling and
  Analysis of Spatio-Temporal Processes in Biomolecular Systems.}}
\newblock In: {\sl \bibinfo{booktitle}{APCCM}}, ~\bibinfo{volume}{96}.
  \bibinfo{publisher}{Australian Computer Society. CPRIT}, pp.
  \bibinfo{pages}{39--48}.
\bibitemend

\bibitemstart{Smart:1998}
\bibinfo{author}{J.~L. Smart} \& \bibinfo{author}{J.~A. McCammon}
  (\bibinfo{year}{1998}): \emph{\bibinfo{title}{Analysis of synaptic
  transmission in the neuromuscular junction using a continuum finite element
  model.}}
\newblock {\sl \bibinfo{journal}{Biophys. J}} \bibinfo{volume}{75}, pp.
  \bibinfo{pages}{1679--1688}.
\bibitemend

\bibitemstart{Smoluchowski:1917}
\bibinfo{author}{M.~V. Smoluchowski} (\bibinfo{year}{1917}):
  \emph{\bibinfo{title}{Versuch einer mathematischen Theorie der
  Koagulationskinetik kolloider Losungen}}.
\newblock {\sl \bibinfo{journal}{Z. Phys. Chem.}} \bibinfo{volume}{92}, pp.
  \bibinfo{pages}{129--168.}
\bibitemend

\bibitemstart{Song:2004}
\bibinfo{author}{Y~Song}, \bibinfo{author}{Y~Zhang}, \bibinfo{author}{T~Shen},
  \bibinfo{author}{C.~L. Bajaj}, \bibinfo{author}{J.~A. McCammon} \&
  \bibinfo{author}{N.~A. Baker} (\bibinfo{year}{2004}):
  \emph{\bibinfo{title}{Finite Element Solution of the Steady-State
  Smoluchowski Equation for Rate Constant Calculations}}.
\newblock {\sl \bibinfo{journal}{Biophys. J.}} \bibinfo{volume}{86}, pp.
  \bibinfo{pages}{2017--2029}.
\bibitemend

\bibitemstart{Takahashi:2008}
\bibinfo{author}{K.~Takahashi} (\bibinfo{year}{2008}): \emph{\bibinfo{title}{An
  Exact Brownian Dynamics Method for Cell Simulations}}.
\newblock In: \bibinfo{editor}{Springer Berlin~/ Heidelberg}, editor: {\sl
  \bibinfo{booktitle}{Computational Methods in Systems Biology 6th
  International Conference CMSB 2008, Rostock, Germany, October 12-15, 2008.
  Proceedings}}. p. \bibinfo{pages}{Computational Methods in Systems Biology}.
\bibitemend

\bibitemstart{Unger2009}
\bibinfo{author}{A.~Unger} \& \bibinfo{author}{H.~Schumann}
  (\bibinfo{year}{2009}): \emph{\bibinfo{title}{Visual Support for the
  Understanding of Simulation Processes}}.
\newblock In: {\sl \bibinfo{booktitle}{Proceedings of IEEE Pacific
  Visualization Symposium}}.
\bibitemend

\bibitemstart{Wade:1993}
\bibinfo{author}{R.~C. Wade}, \bibinfo{author}{M.~E. Davis},
  \bibinfo{author}{B.~A. Luty}, \bibinfo{author}{J.~D. Madura} \&
  \bibinfo{author}{J.~A. McCammon.} (\bibinfo{year}{1993}):
  \emph{\bibinfo{title}{Gating of the active site of triose phosphate
  isomerase: Brownian dynamics simulations of flexible peptide loops in the
  enzyme}}.
\newblock {\sl \bibinfo{journal}{Biophys. J}} \bibinfo{volume}{64}, pp.
  \bibinfo{pages}{9--15}.
\bibitemend

\bibliographyend
\end{thebibliography}

\end{document}